\documentclass[preprint, 5p]{elsarticle}
\usepackage{color}
\usepackage{hyperref}
\begin{document}

\def\anti_particle{\tilde}
\def\cls{CL$_s$ }
\def\cl{\mathrm{CL}}

\begin{frontmatter}

\title{Search for sterile neutrinos at the DANSS experiment}
%\author[mymainaddress,mysecondaryaddress]{Elsevier Inc}

\author[ITEP,MEPhI,MIPT]{I Alekseev}
\author[JINR]{V Belov}
\author[MEPhI,JINR]{V Brudanin}
\author[LPI]{M Danilov\corref{correspondingauthor}}
\cortext[correspondingauthor]{Corresponding author}
\ead{danilov@lebedev.ru}

\author[JINR,DST]{V~Egorov}
\author[JINR]{D~Filosofov}
\author[JINR]{ M~Fomina}
\author[JINR,CTU]{Z~Hons}
\author[JINR,DST]{S~Kazartsev}
\author[ITEP,MIPT]{A~Kobyakin}
\author[JINR]{A~Kuznetsov}
\author[ITEP]{I~Machikhiliyan}
\author[JINR]{D~Medvedev}
\author[ITEP]{V Nesterov}
\author[JINR]{A~Olshevsky} 
\author[ITEP]{N~Pogorelov}
\author[JINR]{D~Ponomarev}
\author[JINR]{I~Rozova}
\author[JINR]{N~Rumyantseva}
\author[ITEP]{V~Rusinov}
%\author[JINR]{A~Salamatin}
\author[ITEP,MEPhI]{E Samigullin}
\author[JINR]{Ye~Shevchik}
\author[JINR]{M~Shirchenko}
\author[JINR,ICL]{Yu~Shitov}
\author[ITEP,MIPT,LPI]{N~Skrobova}
\author[ITEP]{A~Starostin}
\author[ITEP]{D~Svirida}
\author[ITEP]{E~Tarkovsky}
%\author[ITEP]{I~Tikhomirov}
\author[JINR,CTU]{J~Vl\'{a}\v{s}ek} 
%\author[MEPhI]{E~Volkova}
\author[JINR]{I~Zhitnikov}
\author[JINR]{D~Zinatulina}

\address[ITEP]{Alikhanov Institute for Theoretical and Experimental Physics NRC "Kurchatov Institute",\\ B. Cheremushkinskaya 25, Moscow, 117218, Russia}
\address[MEPhI]{National Research Nuclear University MEPhI (Moscow Engineering Physics Institute), Kashirskoe highway 31, Moscow, 115409, Russia}
\address[MIPT]{Moscow Institute of Physics and Technology, 9 Institutskiy per., Dolgoprudny, Moscow Region, 141701, Russia}
\address[JINR]{Joint Institute for Nuclear Research, Joliot-Curie 6, Dubna, Moscow region, 141980, Russia}
\address[LPI]{Lebedev Physical Institute of the Russian Academy of Sciences, 53 Leninskiy Prospekt, Moscow, 119991, Russia}
\address[DST]{Dubna State University, Universitetskaya 19, Dubna, Moscow Region, 141982, Russia}
%\address[NPI]{Nuclear Physics Institute, \v{R}e\v{z} 130, 250 68 \v{R}e\v{z}, Cz, Czech Republic}
\address[CTU]{Czech Technical University in Prague, Zikova 1903/4, 166 36 Prague 6, Czechia}
\address[ICL]{Imperial College London, South Kensington Campus, SW7 2AZ, London, United Kingdom}

\begin{abstract}
DANSS is a highly segmented 1~m${}^3$ plastic scintillator detector. Its 2500 one meter long scintillator strips have a Gd-loaded reflective cover. The DANSS detector is placed under an industrial 3.1~$\mathrm{GW_{th}}$ reactor of the Kalinin Nuclear Power Plant 350~km NW from Moscow. The distance to the core is varied on-line from 10.7~m to 12.7~m. The reactor building provides about 50~m water-equivalent shielding against the cosmic background. DANSS detects almost 5000 $\anti_particle\nu_e$ per day at the closest position with the cosmic background less than 3\%.  The inverse beta decay process is used to detect $\anti_particle\nu_e$. Sterile neutrinos are searched for assuming the $4\nu$ model (3 active and 1 sterile $\nu$). The exclusion area in the $\Delta m_{14}^2,\sin^22\theta_{14}$ plane is obtained using a ratio of positron energy spectra collected at different distances. Therefore results do not depend on the shape and normalization of the reactor $\anti_particle\nu_e$ spectrum, as well as on the detector efficiency. Results are based on 966 thousand antineutrino events collected at three different distances from the reactor core. The excluded area covers a wide range of the sterile neutrino parameters up to $\sin^22\theta_{14}<0.01$ in the most sensitive region. 
%\bigskip
\bigbreak
{\noindent
Published in the Phys.Lett.B as \href{https://doi.org/10.1016/j.physletb.2018.10.038}{doi.org/10.1016/j.physletb.2018.10.038}  
}
\end{abstract}


\end{frontmatter}

%\linenumbers
\section{Introduction}

Oscillations of the three neutrino flavors are well established. Two mass differences and three angles describing such oscillations have been measured \cite{Fogli}. Additional light active neutrinos are excluded by the measurements of the Z boson decay width \cite{PDG}. Nevertheless, existence of additional sterile neutrinos is not excluded. Moreover, indications of several effects observed with about $3\sigma$ significance level can be explained by active-sterile neutrino oscillations. The GALEX and SAGE Gallium experiments performed calibrations with radioactive sources and reported the ratio of numbers of observed to predicted events of $0.88\pm 0.05$ \cite{SAGE}. This deficit is the so called ``Gallium anomaly'' (GA) \cite{GA}. Mueller et al. \cite{Mueller} made new estimates of the reactor $\anti_particle\nu_e$ flux which is about 6\% higher than experimental measurements at small distances. This deficit is the so called ``Reactor antineutrino anomaly'' (RAA). Both anomalies can be explained by active-sterile neutrino oscillations at very short distances requiring a mass-squared difference of the order of 1~eV$^2$ \cite{Mention2011}. The LSND collaboration reported observation of $\anti_particle\nu_\mu \rightarrow \anti_particle\nu_e$ mixing with the mass-squared difference bigger than $\sim 0.1~$eV$^2$ \cite{LSND}. However, results of the MiniBooNE tests of this signal are inconclusive and probably indicate additional effects \cite{MiniBooNE}. 
There are also cosmological constraints on the effective number of neutrinos  \cite{Planck}.
However, in several theoretical models sterile neutrinos are still compatible with these constraints. Details can be found in a review of sterile neutrinos \cite{WhitePaper}.

The survival probability of a reactor $\anti_particle\nu_e$ at short distances in the 4$\nu$ mixing scenario (3 active and 1 sterile neutrino) is described by a familiar expression

\begin{equation}
1-\sin^22\theta_{14}\sin^2\left(\frac {1.27\Delta m_{14}^2 [\mathrm{eV}^2] L[\mathrm m]}{E_\nu [\mathrm{MeV}]}\right).
\end{equation}

The existence of sterile neutrinos would manifest itself in distortions of the $\anti_particle \nu_e$ energy spectrum at short distances. At longer distances these distortions are smeared out and only the rate is reduced by a factor of $1-\sin^2(2\theta_{14})/2$. Measurements at only one distance from a reactor core are not sufficient since the theoretical description of the $\anti_particle \nu_e$ energy distribution is considered not to be reliable enough. The most reliable way to observe such distortions is to measure the $\anti_particle \nu_e$ spectrum with the same detector at different distances. In this case, the shape and normalization  of the $\anti_particle \nu_e$ spectrum as well as the detector efficiency are canceled out. Detector positions should be changed frequently enough in order to cancel out time variations of the detector and reactor parameters. The DANSS experiment uses this strategy and measures $\anti_particle \nu_e$ spectra at 3 distances from the reactor core centre: 10.7~m, 11.7~m, and 12.7~m to the detector centre. 
The detector positions are changed typically 3 times a week. Antineutrinos are detected by means of the Inverse Beta Decay (IBD) reaction
\begin{equation}
\label{eq1}
\anti_particle{\nu}_e + p \rightarrow e^+ + n ~\mbox{with}~ E_{\anti_particle\nu} = E_{e^+} + 1.80~\mathrm{MeV}.
\end{equation}
% \textrm {with E_{\anti_particle\nu} = E_{e^+} + 1.80 MeV}.

\section{The DANSS Detector}
The DANSS detector is described elsewhere \cite{DANSS}. Here we mention only a few essential features.
The DANSS spectrometer does not contain any flammable or other dangerous liquids and may therefore be located very close to the core of a 3.1~GW$_{\rm th}$  industrial power reactor at the Kalinin Nuclear Power Plant (KNPP) 350 km NW of Moscow. DANSS is installed under the reactor core. The reactor cauldron, cooling pond, concrete and other materials provide a good shielding equivalent to $\sim$~50~m of water, which removes the hadronic component of the cosmic background and reduces the cosmic muon flux by a factor of 6. The size of the reactor core is quite big (3.7 m in height and 3.2 m in diameter) which leads to the smearing of the oscillation pattern. This drawback is compensated by a high $\anti_particle{\nu_e}$  flux of $\sim5\times10^{13}\; \anti_particle\nu_e /{\rm cm}^2/{\rm s}$ at a distance of 11~m. 

\begin{figure}[th]%1
%\vspace{-4.2cm}
\centering
\includegraphics[width=0.95\linewidth]{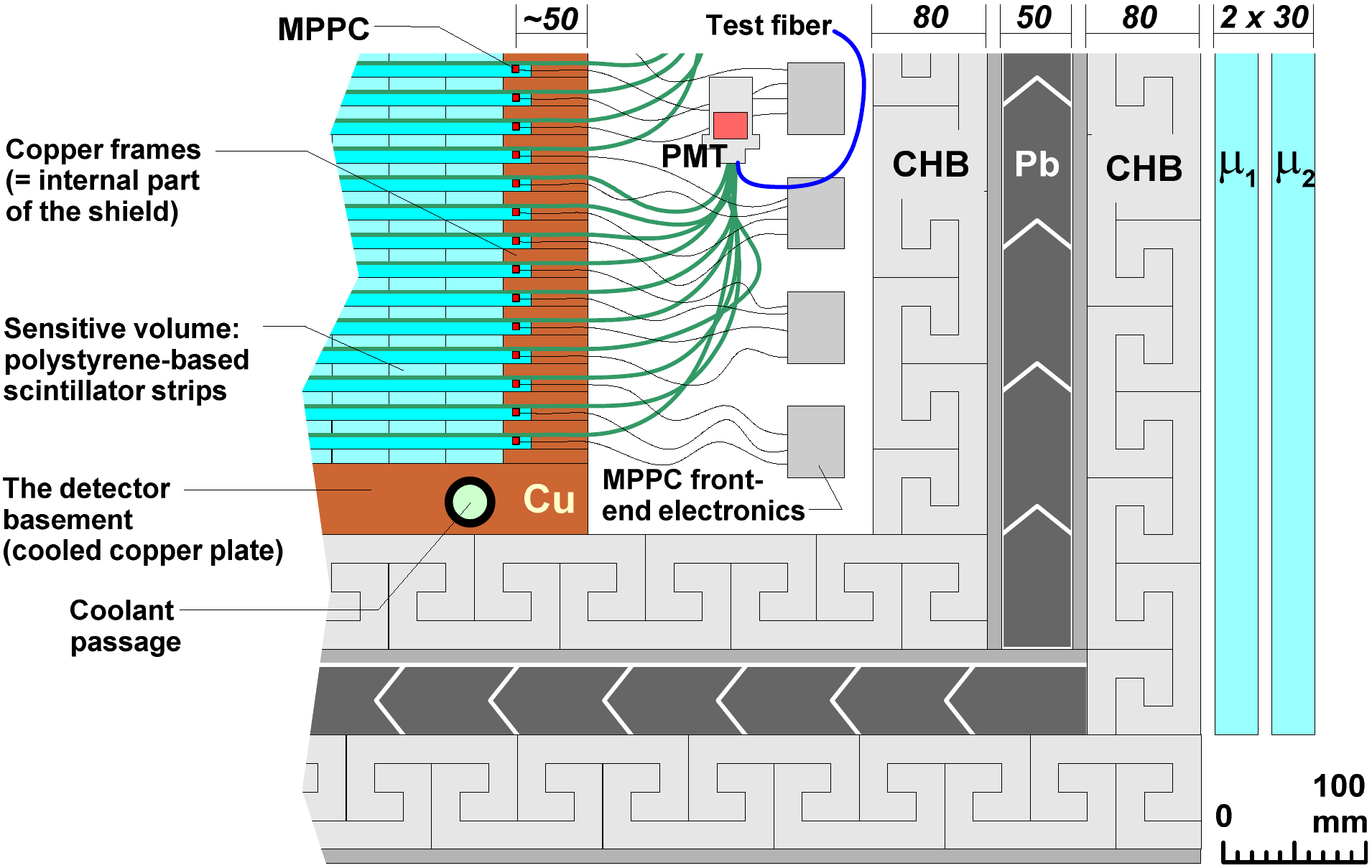}
 \caption{\footnotesize Simplified cross section of a corner of the DANSS detector.}
 \label{DANSS}
\end{figure}

DANSS is a highly segmented plastic scintillator detector with a total volume of 1~m$^3$, surrounded with a composite shield of copper (Cu --- 5~cm), borated polyethylene (CHB --- 8~cm), lead (Pb --- 5~cm) and one more layer of borated polyethylene (CHB --- 8~cm) (see Fig.~\ref{DANSS}). It is surrounded on 5 sides (excluding bottom) by double layers of 3~cm thick scintillator plates to veto cosmic muons.

The basic element of DANSS is a polystyrene-based extruded scintillator strip ($1 \times 4 \times 100$~cm${}^3$) with a thin ($\sim 0.2$~mm) Gd-containing surface coating. The amount of Gd in the detector is 0.35$\%_{wt}$. The coating serves as a light reflector and a ($n,\gamma$)-converter simultaneously. 

Light from the strip is collected with three wavelength-shifting (WLS) Kuraray fibers Y-11, $\oslash$~1.2~mm, glued into grooves along the strip. One (blind) end of each fiber is polished and covered with a mirror paint, which decreases the total longitudinal attenuation of a light signal to about $30\%/$m. This non-uniformity can be corrected using information from orthogonal strips or from the neutron capture position, which is typically $\sim~10$~cm away from the positron production point. The response non-uniformity across the strip is $\sigma =7.8\% $\cite{nonuniformity}. This non-uniformity of the response can't be corrected. It leads, together with the energy losses in the inactive strip reflective layers, to the deterioration of the energy resolution in comparison with naive estimates from the photoelectron statistics. These effects are included into the MC simulation of the detector.

\begin{figure}[th]%2
%\vspace{-4.2cm}
\centering
\includegraphics[width=0.95\linewidth]{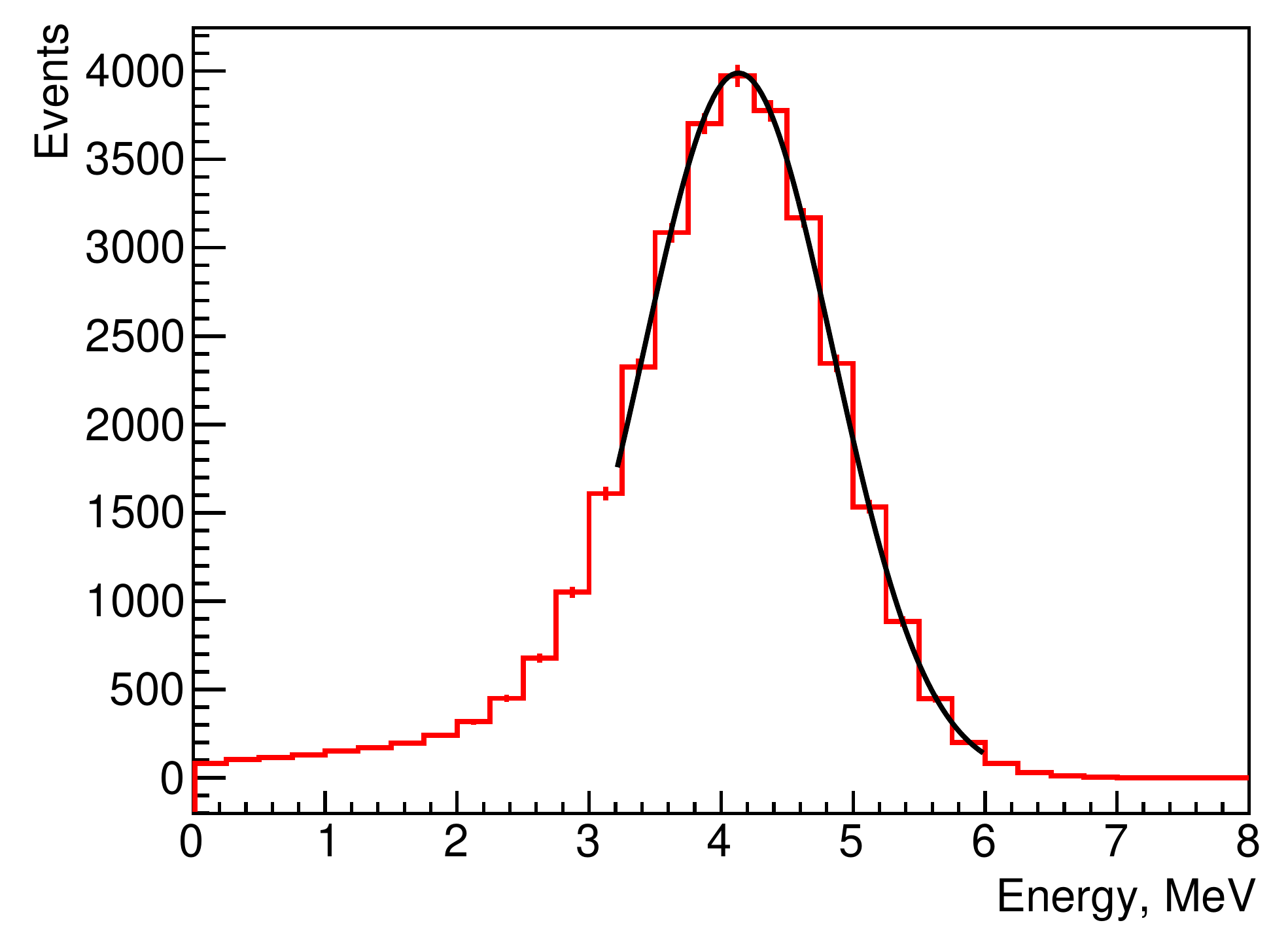}
 \caption{\footnotesize Reconstructed positron energy for 4.125~MeV positrons from MC simulations. The curve represents the fit of the Gaussian part of the spectrum.}
 \label{resolution4MeV}
\end{figure}

Each 50 parallel strips are combined into a module, so that the whole detector (2500 strips) is a structure of 50 intercrossing modules. 
Each module is viewed with a compact photomultiplier tube (PMT) (Hamamatsu R7600U-300) coupled to all 50 strips of the module via 100 WLS fibers, two per strip. 
PMTs are placed inside the shielding but outside the copper layer, which serves also as a module frame.
In addition, to get a more precise energy and space pattern of an event, each strip is equipped with an individual Silicon PhotoMultiplier (SiPM) (MPPC S12825-050C(X)) coupled to the strip via the third WLS fiber. The SiPM is fixed directly at the end of the strip using a plastic light connector.
All signals are digitized with specially designed 12~bit, 125~MHz FADCs\cite{FADC}. Only front-end electronics is placed inside the shielding but outside the Cu layer. All other electronics is placed outside of the detector shielding. One 6U VME board serves 64 channels. 
 SiPMs (PMTs) register about 18 (20) photo-electrons (p.e.) per MeV. These numbers were obtained using measurements with cosmic muons and artificially driven LEDs. So the total number is 38 p.e./MeV. 
Parameterized strip response non-uniformities have been incorporated into the GEANT4 (Version 4.10.4) MC simulation of the detector. The MC simulation included also a spread in the light yields of different strips, dead channels, Poisson fluctuations in the number of p.e. at the first 2 PMT dinodes, the excess noise factor for SiPMs due to the optical cross-talk between pixels. The experimental energy resolution for cosmic muon signals in the scintillator strips is 15\% worse than that from the MC calculation.
Therefore, the MC estimations are scaled up by the corresponding factor. Figure~\ref{resolution4MeV} shows the simulated DANSS response to a 4.125 MeV positron signal. The energy resolution is modest ($\sigma/E=17\%$ for the Gaussian part of the spectrum). This leads to additional smearing of the oscillation pattern, comparable with the smearing due to the large reactor core size.

\begin{figure}[th]%3
%\vspace{-5.5cm}
\centering
\includegraphics[width=0.95\linewidth]{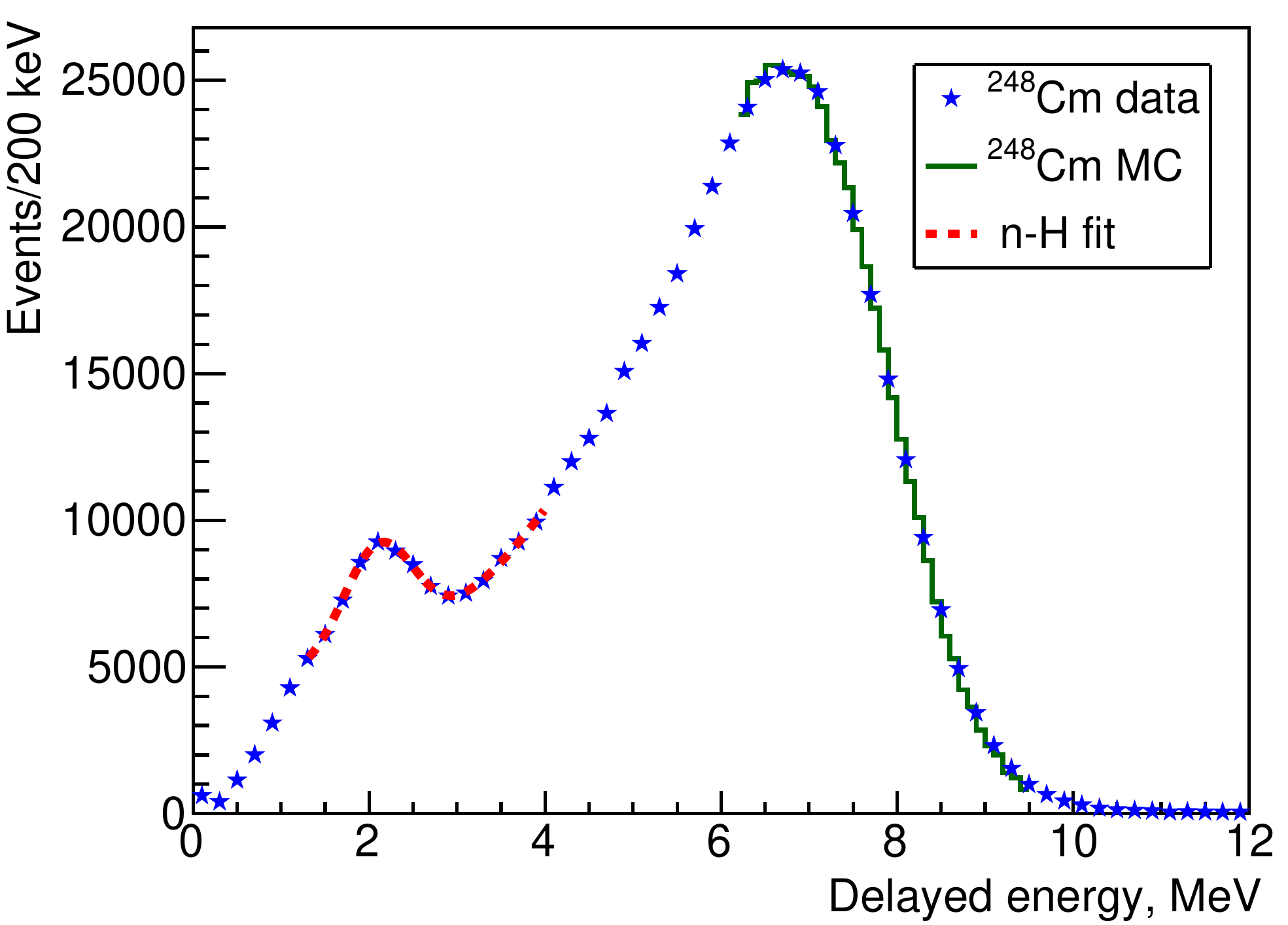}
 \caption{\footnotesize Energy spectrum of the delayed signals measured with the $^{248}$Cm neutron source. The dashed curve shows the fit of the n-H peak. The histogram is the MC prediction for the n-Gd peak}
 \label{Cm}
\end{figure}

\begin{figure}[th]%4
\centering
\includegraphics[width=0.95\linewidth]{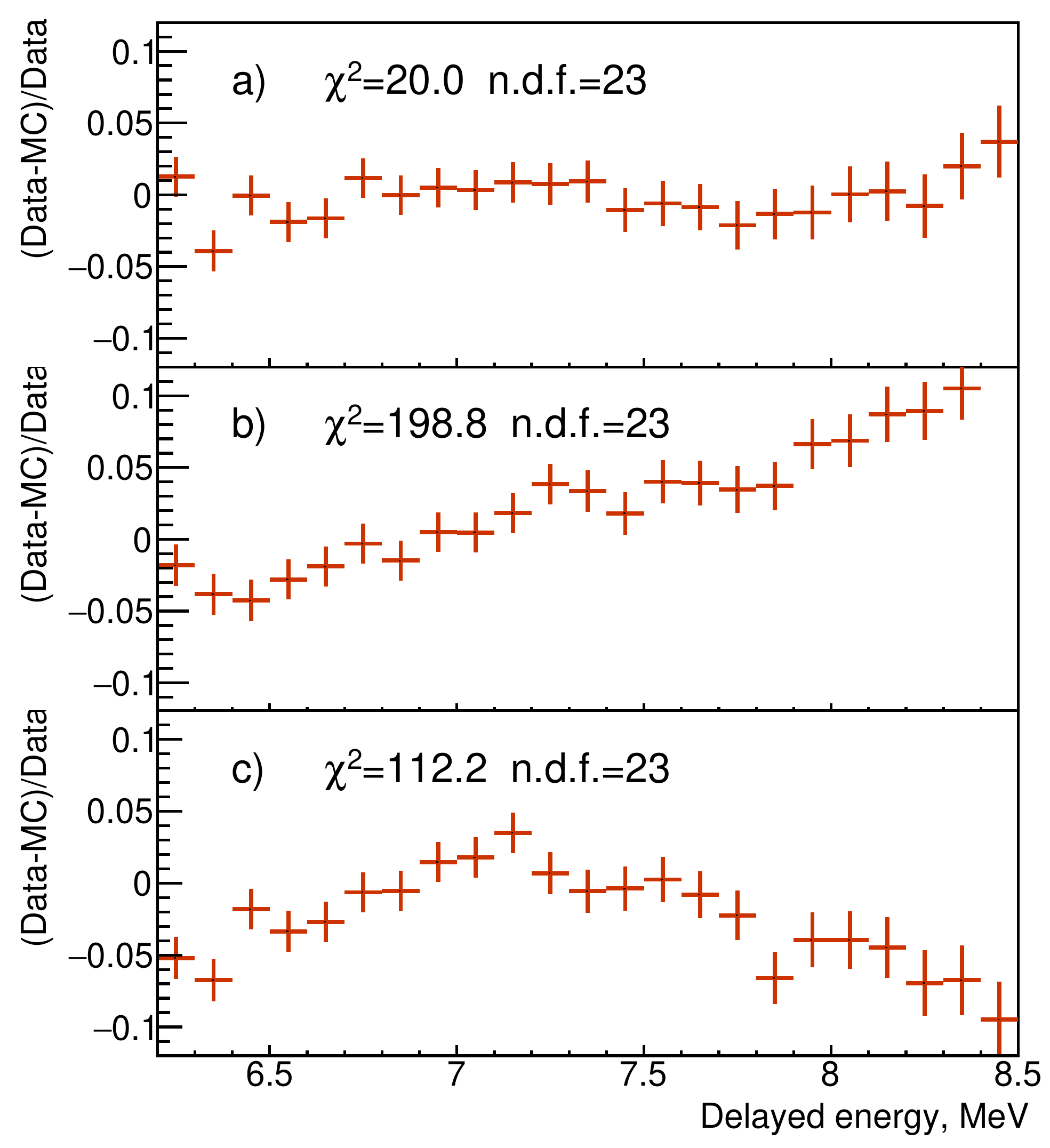}
 \caption{\footnotesize Ratio of the difference in MC predictions and data divided by the data for the neutron capture signal on Gd for:
a) Nominal MC parameters;
b) The MC energy scale increased by 1\%;
c) The MC energy resolution worsened by 5\% of its nominal value.
$\chi^2$ values correspond to the deviations of the points from the zero line.}
 \label{Gd}
\end{figure}

Figure~\ref{Cm} shows the energy distribution of neutron capture signals from a $^{248}$Cm source placed at the center of the detector. Two peaks correspond to the neutron capture by protons and by Gd. The fit of the first peak gives a resolution compatible with the MC simulations (see Table ~\ref{1}). The MC describes well the high energy part of the n-Gd peak, although there is some tension in the tail. Figure~\ref{Gd}(a) shows the difference between the MC and data divided by the data for the main part of the n-Gd peak. There is a good agreement between the MC and data. On the other hand, the relative changes of 1\% in the MC positron energy scale or of 5\% in the MC energy resolution lead to serious discrepancies between the data and MC (see Fig.~\ref{Gd}(b) and Fig.~\ref{Gd}(c)).  This comparison provides the upper limits on the systematic uncertainties of these parameters. It is very difficult to simulate reliably the low energy part of the n-Gd capture signal since there are many cascade decay chains with unknown probabilities. Our MC does not describe well the lower energy part of the spectrum and hence we do not use it for the comparison.

\begin{figure}[th]%5
\centering
\includegraphics[width=0.95\linewidth]{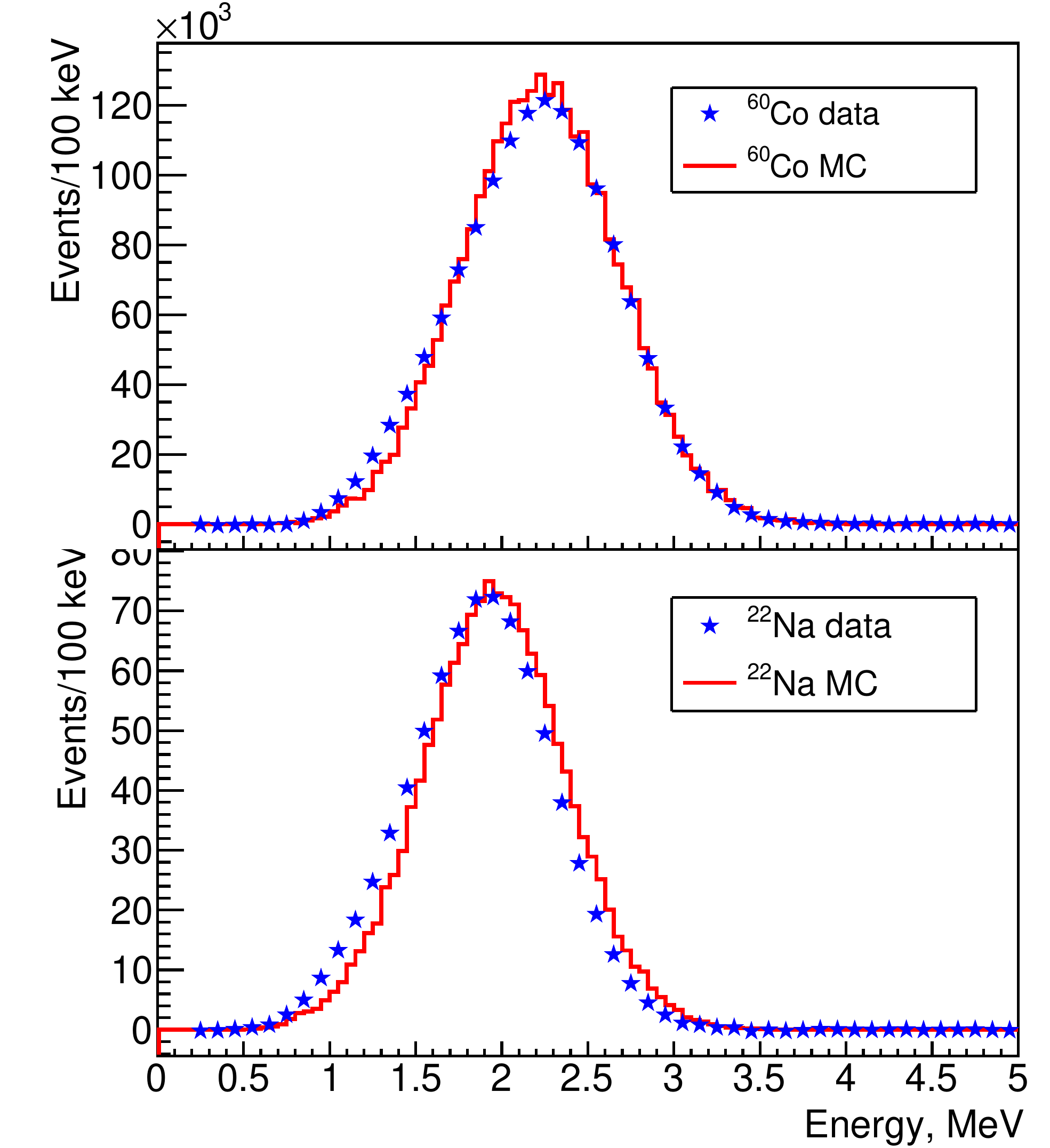}
 \caption{\footnotesize Detector response to the $^{60}$Co and $^{22}$Na sources. Histograms are the MC predictions.}
 \label{Co}
\end{figure}

%(Add discussion of the Gd fit)
Figure~\ref{Co} shows the energy distribution of signals from a $^{60}$Co and $^{22}$Na $\gamma$-sources placed in the center of the detector. The observed energy resolutions and peak positions are consistent with the MC expectations (see Table~\ref{1}). The visible energy in the detector from the radioactive sources is lower than the source energy due to losses in inactive strip reflective layers, dead channels and photons escaped from the detector.   

\begin{table}[th]
 \caption{\footnotesize Comparison of the data and MC results for different radioactive sources. Peak energies and $\sigma$ are given in MeV. Statistical errors of the fit results are negligible. In case of the n-Gd peak the effective description of the sum of two isotopes with one Gaussian is presented.}
 \label{1}
%\centering
\begin{tabular}{|c|cccc|}
\hline
 & E, data & E, MC & $\sigma$, data & $\sigma$, MC\\
 \hline
$^{22}$Na & 1.90 & 1.96 & 0.40 & 0.42\\
$^{60}$Co & 2.22 & 2.22 & 0.46 & 0.45\\
$^{248}$Cm -- H(n,$\gamma$) & 2.04 & 1.97 & 0.49 & 0.49\\
$^{248}$Cm -- Gd(n,$\gamma$) & 6.76 & 6.80 & 1.08 & 1.03\\
 \hline
\end{tabular}
\end{table}

\begin{figure}[th]%6
\centering
\includegraphics[width=0.95\linewidth]{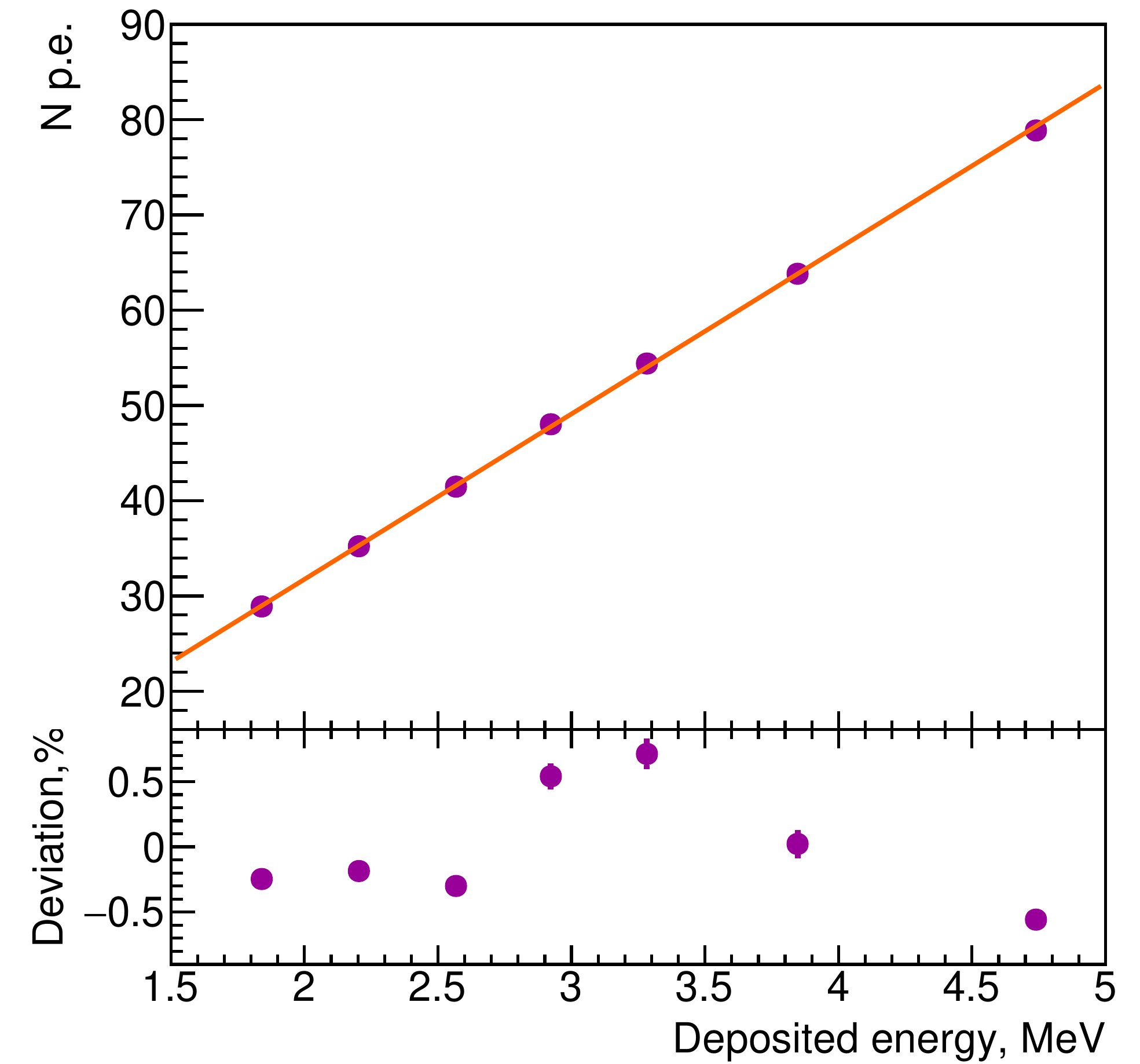}
 \caption{\footnotesize Number of p.e. detected with SiPMs vs energy deposited by cosmic muons in a scintillator strip (top) and the deviation from a straight line fit (bottom). Different muon energies correspond to different crossing angles of the muons and the strips.}
 \label{Calibration}
\end{figure}

SiPM gain calibration was performed using noise signals typically every 5 days. Calibration with cosmic muons of all strips in the whole detector was also performed once in $\sim 5$ days. A detailed description of the calibration procedure is presented elsewhere \cite{Calibration}.
Using the high granularity of the detector, we reconstruct muons crossing the strips at different angles. For each angle, we calculate the track length in the strip and the deposited energy in p.e. The same procedure is performed using MC muons but in this case, the corresponding energy deposition in MeV is obtained from the GEANT4 simulations. For each bin in the muon crossing angle the most probable energy values are obtained for the data and MC using the fits of the corresponding Landau spectra.  The measured energy versus the calculated energy is shown in Fig. ~\ref{Calibration}. The response is linear within 0.7\% in the (1.8---4.7)~MeV energy range.
The energy measured with PMTs is proportional on average to the energy, measured with SiPMs. This was checked comparing the positron energy from IBD events reconstructed with SiPMs and with PMTs. The PMT linearity was also checked using the LED calibration system. Therefore, the PMT energy response is also linear. Positrons with energies higher than 4.7~MeV typically deposit their energy in several strips. Therefore, the detector response should be linear for high energies as well.

\section{Data taking and analysis}

\label{data_analysis}

\begin{figure}[h]%7
\begin{minipage}{0.47\textwidth}
\includegraphics[width=\textwidth]{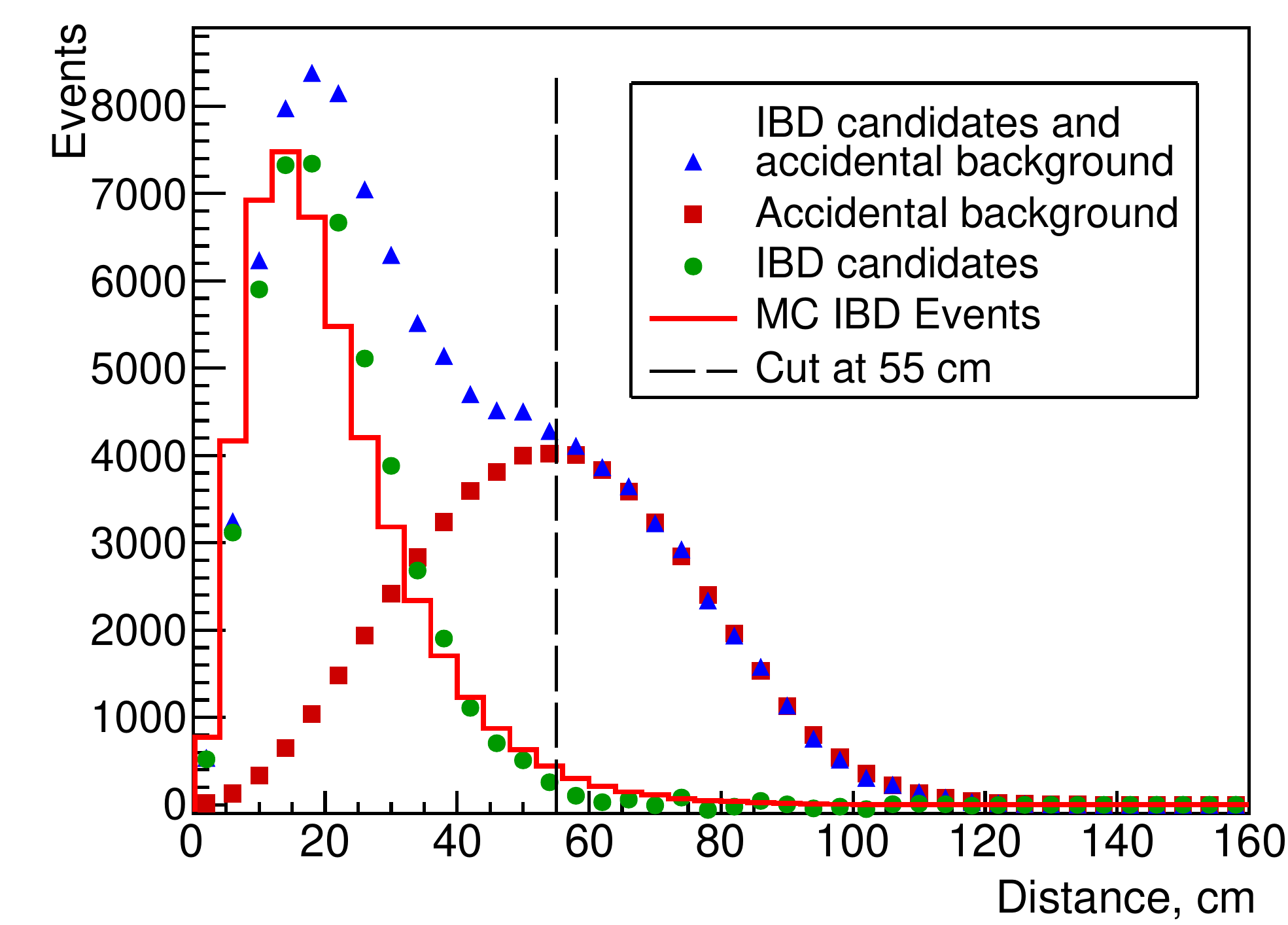}
\caption{\label{Distance}Distance between positron and neutron reconstructed positions. Errors are smaller than sizes of points. The histogram is the MC prediction for the IBD events.}
\end{minipage}\hspace{0.06\textwidth}
\begin{minipage}{0.47\textwidth}%8
\includegraphics[width=\textwidth]{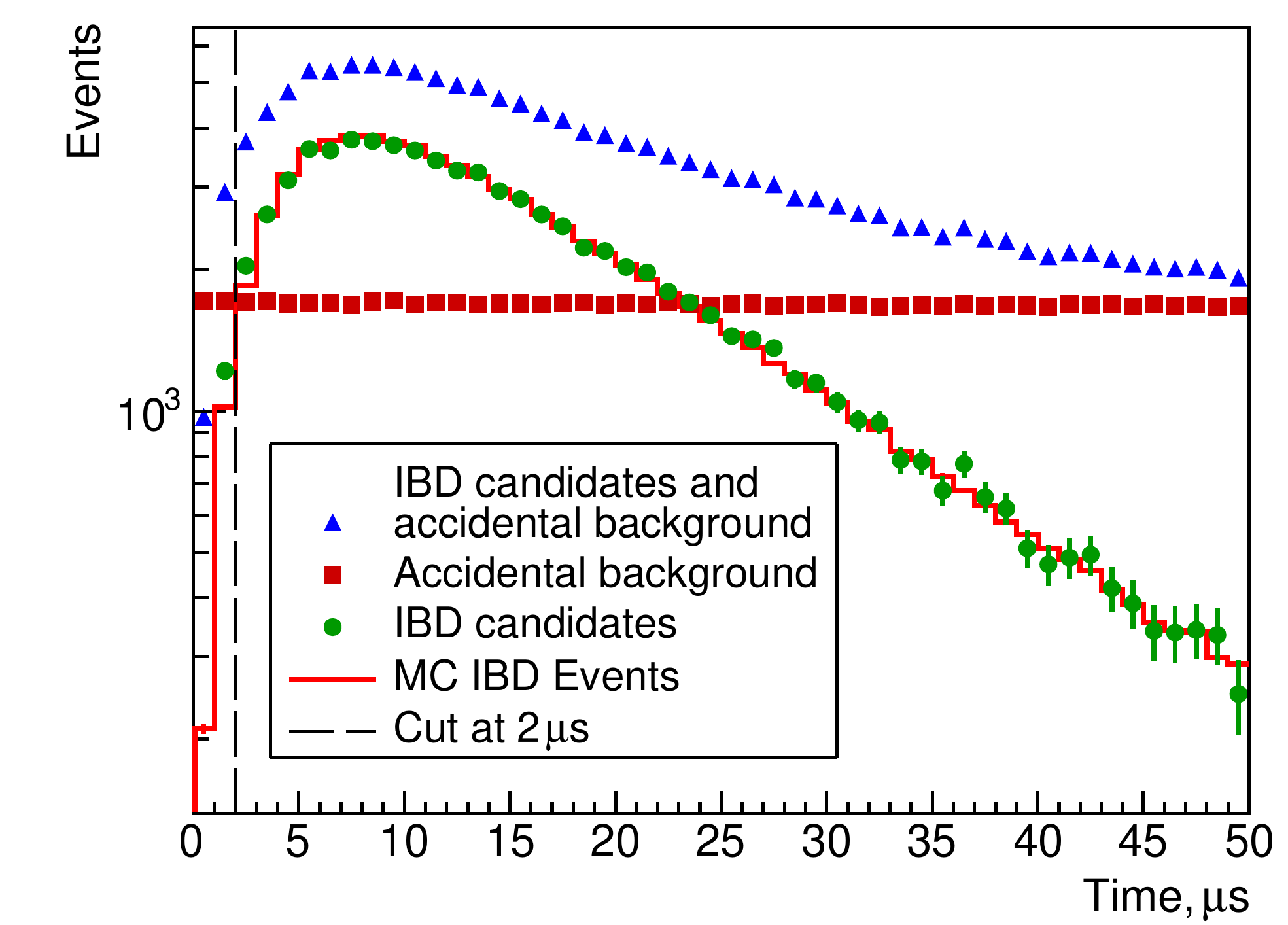}
\caption{\label{Time}Time between prompt and delayed signals. Errors are smaller than sizes of points.The histogram is the MC prediction for the IBD events.}
\end{minipage}
\end{figure}

The trigger of the experiment is produced when the digital sum of all PMT signals is above 0.7~MeV or the energy in the veto system is larger than 4~MeV. The IBD process appears in the data as two distinct events, prompt and delayed.  For each trigger, waveforms for all SiPMs and PMTs are recorded in 512~ns windows. 
The data analysis is performed in several stages. The first one is the noise cleanup. At this step an
average event time is calculated and all hits in SiPMs more than 15~ns away are rejected.  Then single pixel SiPM
hits which have no confirmation by the corresponding PMT are also rejected. After these 
two steps less than 2 noise SiPM hits per event survive out of about 40 in the original sample.
At the next stage various characteristics of the event such as the total energy, the number of hits 
and so on are calculated. At this stage we also look for a continuous ionization cluster and 
calculate its visible energy. The visible energy of a positron cluster is converted using MC simulations into the deposited energy by taking into account average losses in the inactive reflective layers of the strips and dead channels. Sometimes photons from the positron annihilation produce signals in the strips attributed to the positron cluster. This leads to an increase of the visible energy. Such a shift is also corrected on average using MC simulations.
A typical size of the total correction is $\sim2\%$.   
The next step is a search for the time-correlated pairs of prompt-delayed events.
We start with searching for an event with more than 3.5~MeV energy deposit. This is a delayed event candidate unless it has the muon veto (more than 4~MeV energy in a veto counter or more than 1 hit in the veto counters with a low threshold of $\sim$0.15~MeV or more than 20~MeV energy in the main detector). 
Then we look backward in time searching for a prompt event with more than 1~MeV in the positron cluster and no muon veto.
An IBD candidate pair is considered found if the time difference between the prompt and delayed events is in the range (2---50)~$\mu$s.
For a valid pair we also require no event with the muon veto within 60~$\mu$s before the prompt signal (within 200~$\mu$s if E~$>$~300~MeV is released in the main detector).  No other event should occur within 45~$\mu$s before
and 80~$\mu$s after the prompt event.
The found pairs of prompt and delayed events form the experimental sample of IBD candidates.
The accidental background is a random coincidence of prompt (positron-like) and delayed (neutron-like) signals. It is proportional to the product of such signal rates. It does not depend on the difference in time between the corresponding signals since they are not correlated (the backgrounds from correlated pairs of signals are determined separately). Therefore, the accidental background can be directly measured in a model independent way by measuring the coincidence rate of electron-like and neutron-like signals sufficiently separated in time to exclude any correlation. This can be done for many independent time windows. Therefore, the accidental background can be measured with negligible statistical errors. Similar to the experimental sample of IBD candidates, the accidental coincidence sample is formed by looking for a prompt signal in 16 regions: 5, 10,..., 80 ms before the neutron candidate. This sample provides us with a model-independent measure of the accidental background with negligible statistical errors. Distributions for IBD candidates with accidental background, the accidental background and their difference, which represents the IBD candidate events without accidental background are presented in figures~\ref{Distance} and \ref{Time} for about 10\% of the data taken first.

Figure~\ref{Distance}  shows the distance between positron and neutron candidates. The width of the signal distribution is determined  by the neutron travel in the detector before the capture and the uncertainty of the neutron capture vertex reconstruction from the observed $\gamma$-flash. The distribution of the neutron capture time from the positron production moment is presented in Fig.~\ref{Time}. The MC predictions agree nicely with the data.
  These two figures illustrate also the method used to get signal distributions:
a parameter is plotted for the experimental data set of the IBD candidates and for the accidental background data set using the same selection criteria. A signal distribution is obtained as a difference of these two distributions. 
It is worth mentioning that the very procedure of the accidental background subtraction only negligibly increases the statistical errors in the signal distributions. This is because the accidental background is calculated using the 16 times larger sample of pairs. However, the final errors are determined by the initial counts in the IBD candidate sample, which include the accidental background.
Therefore, several cuts are applied, in order to reduce the accidental background. They also reduce slightly the background from neutrons produced by cosmic muons. The cuts are designed to be very soft with respect to the signal in order to avoid any distortions. All cuts were selected without looking at the final results. They have been fixed after collection of about 10\% of the data. The cuts include the following requirements:
\begin{itemize}
\item The distance between positron and neutron candidates should be shorter than 45~cm (55~cm) for a positron candidate reconstructed in 2 (3) dimensions;
\item The additional energy outside a positron cluster should be less than 1.8~MeV, and the most energetic hit should have the energy less than 0.8~MeV;
\item The number of hits in the delayed event is higher than 3;
\item The number of hits in the prompt event outside the positron cluster is less than 11;
\item The most energetic hit of the positron cluster lies in a fiducial volume of the detector which does not include outer strips in the X and Y directions as well as 4 highest and 4 lowest strips in the vertical Z direction. This cut excludes regions with fast changes of the efficiency. It is also useful against the background from fast neutrons.
\end{itemize}

\begin{figure}[h]%9
\begin{center}
\includegraphics[width=0.45\textwidth]{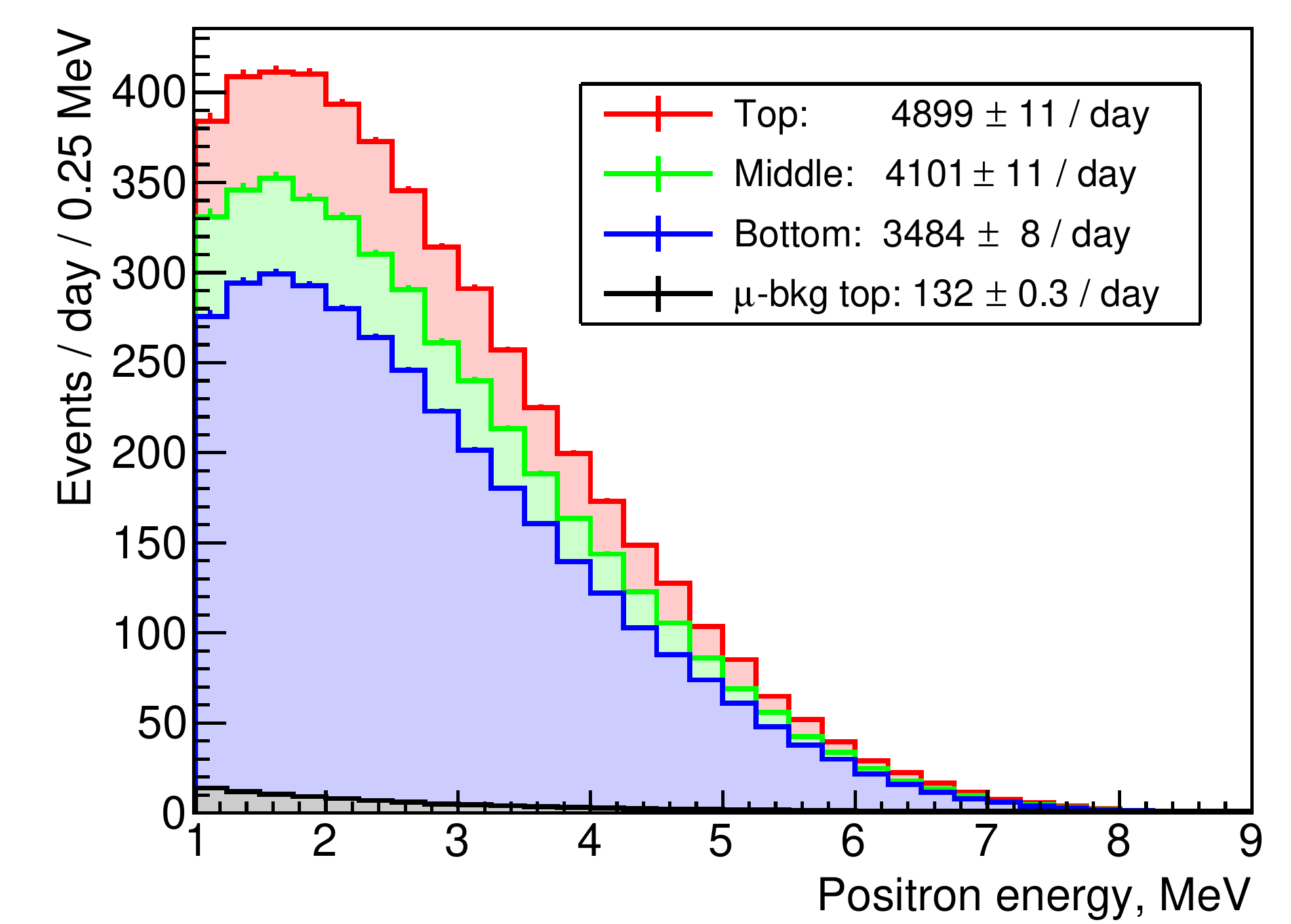}
\end{center}
\caption{\label{Spectra} Positron energy distributions measured at different detector positions after subtraction of all backgrounds 
(statistical errors only). The background from neutrons produced by cosmic muons is shown for the top detector position.}
\end{figure}

Positron energy spectra for 3 detector positions (top, middle, bottom) are shown in Fig.~\ref{Spectra} with
statistical errors only. The corresponding numbers of events are 367, 260 and 339 thousand in the (1---8) MeV energy range.
The rates are not corrected for the detector efficiency which depends on the positron momentum. This efficiency is irrelevant for the present analysis.
The ratios of the positron spectra at the bottom and middle detector positions to the spectrum at the top detector position are summarized in Table~\ref{2}.

The IBD counting rate is 4899 events per day in the top position.
The positron energy does not include annihilation photons and hence is 1.02~MeV lower than the usually used prompt energy.  
The muon-induced background (discussed below) is subtracted.  
Three other reactors at the KNPP are 160~m, 334~m, and 478~m away from the DANSS detector.
The IBD counting rate from these reactors is 0.6\% of the IBD counting rate from the nearest reactor for the top detector position. This contribution is taken into account by the corresponding reduction of the normalization of the obtained spectra.

The energy spectrum of the background from fast neutrons produced outside the detector shielding is estimated by a linear extrapolation from a (10---16)~MeV region to lower energies. 
This background calculation and subtraction is performed separately for the positron candidate energy spectra with and without muon veto. This background constitutes $\sim0.1\%$ of the IBD signal in the (1---8)~MeV region.
%I moved this paragraph up

\begin{figure}[th]%10
\centering
\includegraphics[width=0.95\linewidth]{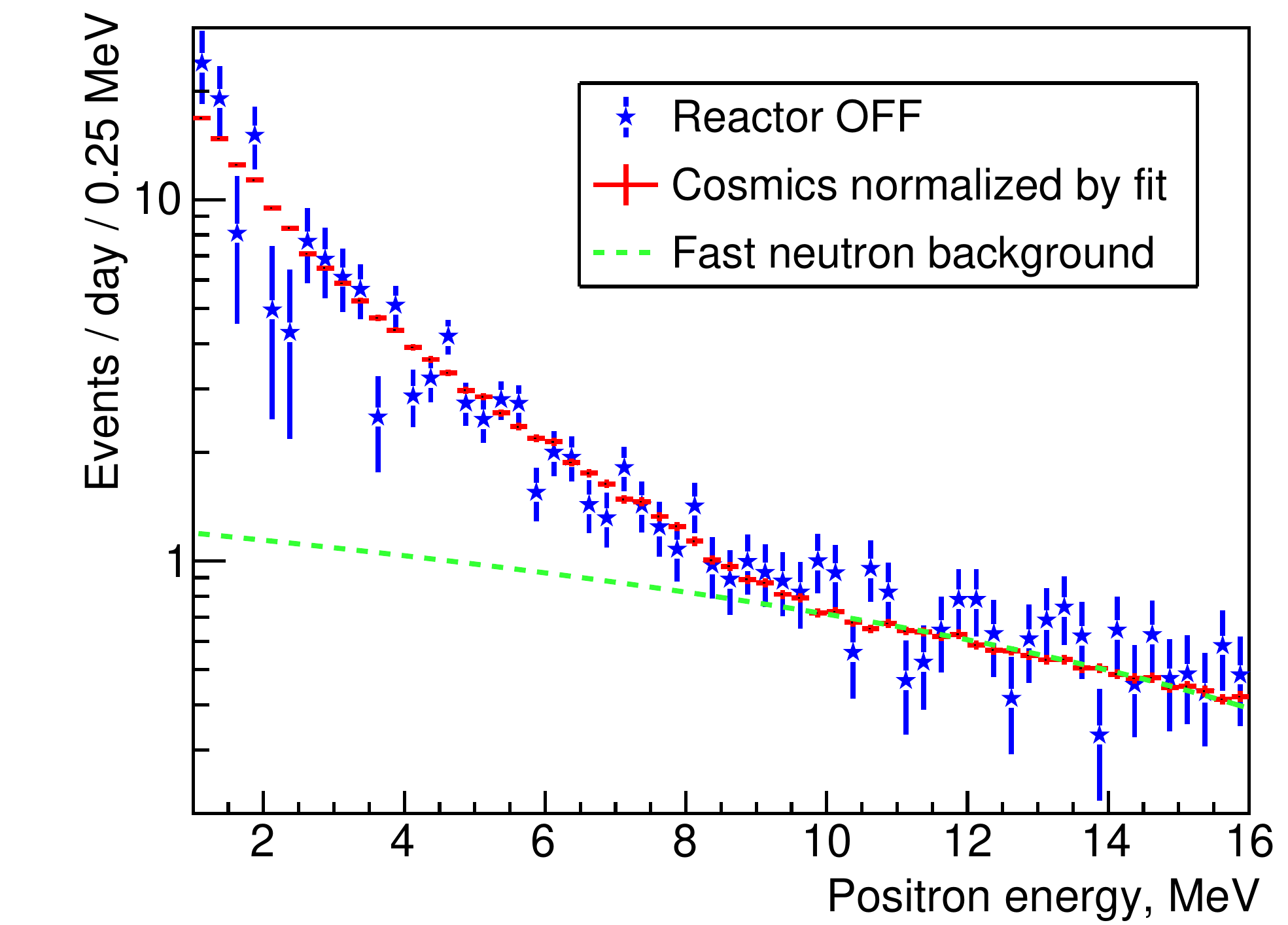}
 \caption{\footnotesize Positron candidate energy distribution during the ``reactor off'' period, statistical errors only. The well measured accidental background and the contribution from the adjacent reactors (29 events/day) are subtracted. The dashed line is the background from fast neutrons produced outside the detector estimated by the extrapolation from the (10---16)~MeV region. Crosses represent a fit of the remaining part of the background using the shape of the background from neutrons produced by muons inside the detector.}
 \label{ReactorOff}
\end{figure}

The energy spectrum of the background from neutrons produced by muons inside the veto system is obtained from events with the muon veto. The amount of this background is determined from a fit of the positron candidate energy spectrum during ``reactor off'' periods using the shape of the background determined from events with the muon veto (see Fig.~\ref{ReactorOff}). This procedure reduces uncertainties in the background shape to a negligible level. A possible small uncertainty in the background rate is taken into account during systematic error studies. This is the most important background. It constitutes 2.7\% of the IBD rate at the top detector position.

A background from $^9$Li and $^8$He produced by cosmic muons was estimated using the time distribution between cosmic events with the energy deposit in the detector bigger than 800~MeV and the IBD candidates. No evidence of the exponential decay component with known $^9$Li decay time of 257.2~ms was found. The corresponding upper limit on this type of background is 5.4 events/day at the 90\% confidence level. 

The shape of the positron spectrum agrees roughly with the MC predictions based on the $\anti_particle\nu_e$ spectrum from \cite{Huber, Mueller}. However, a quantitative comparison requires additional studies of calibration and systematic errors and improvements in the MC simulation of the detector. Since the results of the present analysis practically do not depend on the $\anti_particle\nu_e$ spectrum shape and normalization we postpone these studies till a forthcoming paper. 

\begin{figure}[h]
\begin{center}
\includegraphics[width=0.45\textwidth]{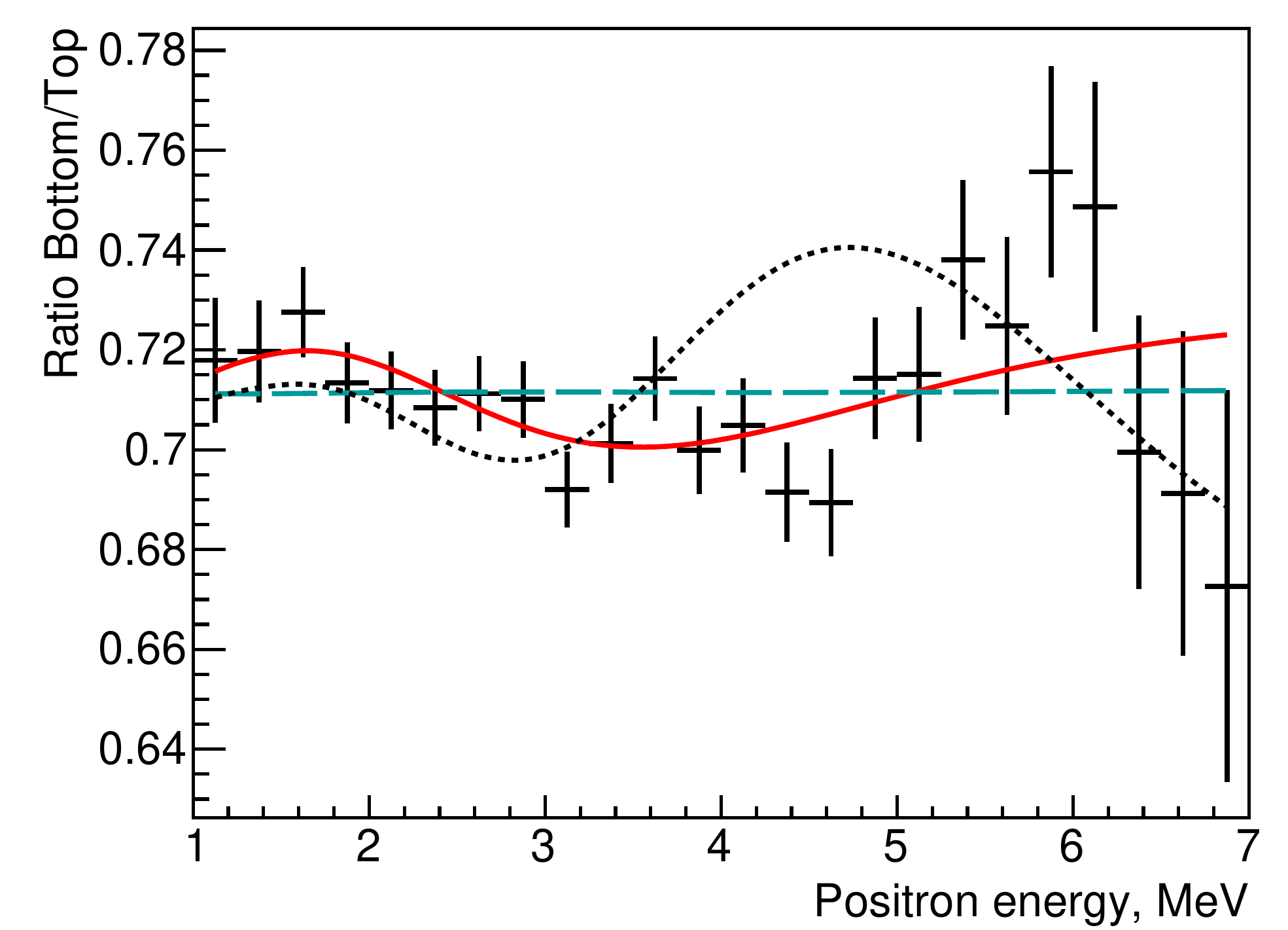}
\end{center}
\caption{\label{Ratio}Ratio of positron energy spectra measured at the bottom and top detector positions (statistical errors only). The dashed curve is the prediction for 3$\nu$ case ($\chi^2=35.0$, 24 degrees of freedom). The solid curve corresponds to the best fit in the $4\nu$ mixing scenario ($\chi^2 = 21.9$, $\sin^22\theta_{14} = 0.05$, $\Delta m_{14}^2 = 1.4~\rm{eV}^2$). The dotted curve is the expectation for the optimum point from the RAA and GA fit \cite{Mention2011} ($\chi^2=83$, $\sin^22\theta_{14}=0.14$, $\Delta m_{14}^2 = 2.3~\rm{eV}^2$)}
\end{figure}

Figure~\ref{Ratio} shows the ratio of positron energy spectra at the bottom and top detector positions. 

\begin{figure}[h]
\begin{center}
\includegraphics[width=0.45\textwidth]{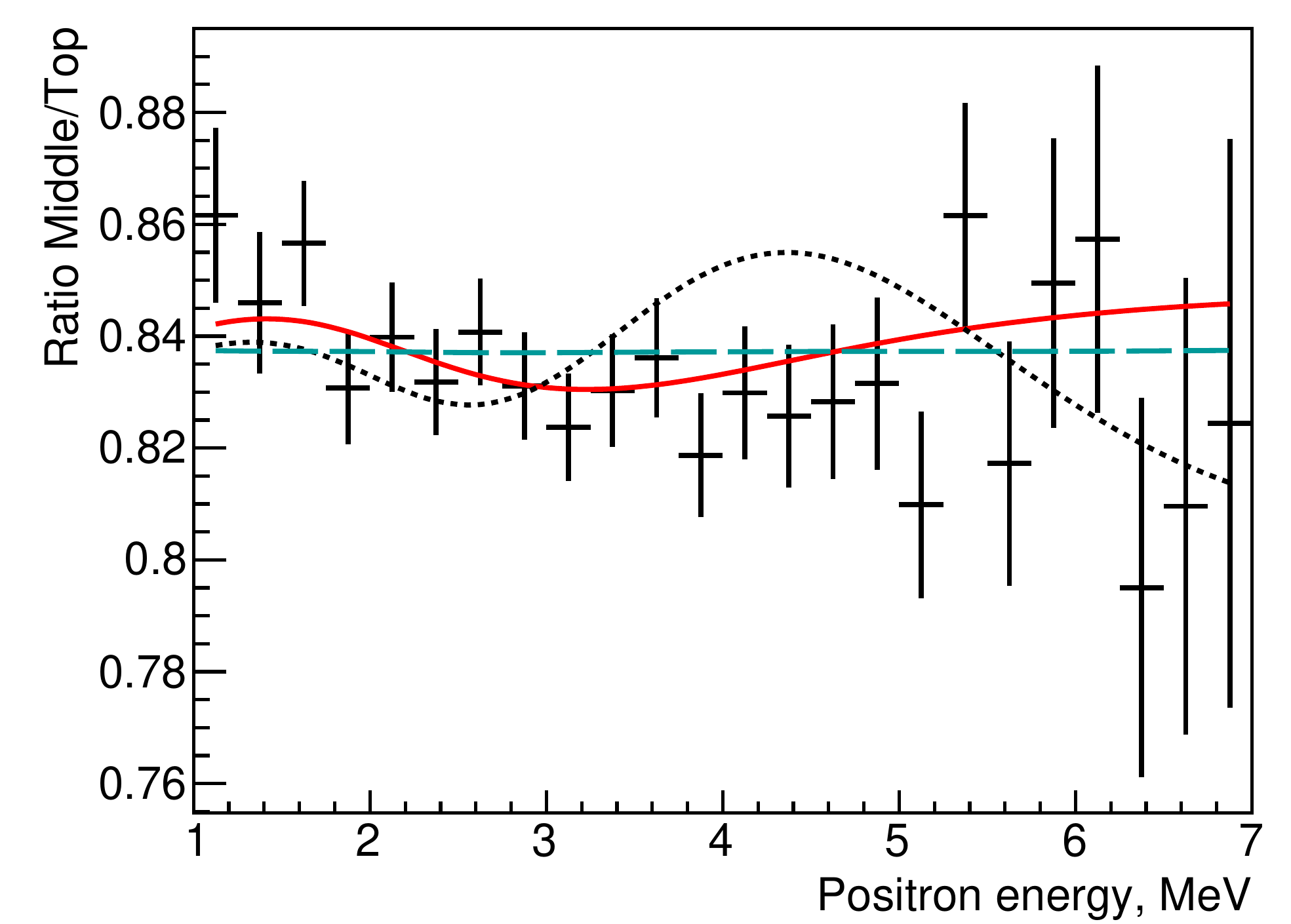}
\end{center}
\caption{\label{RatioMT}Ratio of positron energy spectra measured at the middle and top detector positions (statistical errors only). The dashed curve is the prediction for 3$\nu$ case ($\chi^2 = 21.6$, 24 degrees of freedom). The solid curve corresponds to the best fit in the $4\nu$ mixing scenario for the bottom/top ratio ($\chi^2 = 17.4 $, $ \sin^22\theta_{14} = 0.05$, $\Delta m_{14}^2 = 1.4~\rm{eV}^2$). The dotted curve is the expectation for the optimum point from the RAA and GA fit \cite{Mention2011} ($\chi^2 = 42.7$, $\sin^22\theta_{14}=0.14$, $\Delta m_{14}^2 = 2.3~\rm{eV}^2$)}
\end{figure}

The exclusion area in the sterile neutrino parameter space was calculated using the Gaussian \cls method \cite{CLS} assuming only one type of sterile neutrinos. For a grid of points in the $\Delta m_{14}^2,~\sin^22\theta_{14}$ plane predictions for the ratio $R^{\mathrm{pre}}(E)$  of positron spectra at the bottom and top detector positions were calculated. Calculations included the MC integration over the $\anti_particle\nu_e$  production point in the reactor core,  $\anti_particle\nu_e$  detection point in the detector, and positron energy resolution. The $\anti_particle\nu_e$  production point distributions in the reactor core were provided by the KNPP for different time periods. These distributions were almost flat in the reactor core radius and height with a fast drop near the edges of the core.  The distribution averaged over the campaign was used in the calculations. It was checked that this approximation practically did not influence the final results. The $\anti_particle\nu_e$  energy spectrum from \cite{Huber,Mueller} averaged over the campaign was used for the calculations. However, the final result practically did not depend on the shape and normalization of the $\anti_particle\nu_e$ spectrum since only the ratio of measured positron spectra at the different positions was used in the analysis. This was checked by repeating the analysis using the spectrum from the beginning of the campaign. 

The obtained theoretical prediction for a given point in the $\Delta m_{14}^2,~\sin^22\theta_{14}$ plane was compared with the prediction for the three neutrino case using the Gaussian \cls method for the 95\% and 90\% confidence level (CL) exclusion area estimation. The difference in $\chi^2$ for the two hypotheses $\Delta\chi^2 = \chi^2_{4\nu} - \chi^2_{3\nu}$ was used for the comparison. This difference in $\chi^2$ has a Gaussian distribution with the mean value and width determined from the fit to the Asimov data set, an infinite statistics data sample with parameters fixed to the $4\nu$ or $3\nu$ hypotheses\cite{CLS}. Confidence levels for the $4\nu$ ($\cl_{4\nu}$) and $3\nu$ ($\cl_{3\nu}$)  hypotheses were calculated using the obtained parameters of the Gausssian distributions for these hypotheses and the observed value of $\Delta\chi^2$. These confidence levels provide a measure for the consistency with the $4\nu$ and $3\nu$ hypotheses correspondingly and \cls $= \cl_{4\nu}/\cl_{3\nu}$. The $4\nu$ hypothesis for a given point in the sterile neutrino parameter space is excluded at the 95(90)\% CL if \cls$<1-0.95(0.9)$. 
It was checked that for the majority of the $\Delta m_{14}^2$ values the limits obtained with the \cls method are more conservative than the limits obtained with the raster scan method~\cite{Raster}.
More details on the application of the \cls method in the present analysis are given in Appendix.

The $\chi^2$ for each hypothesis was constructed using 24 data points $R^{\mathrm{obs}}_i$ in the (1---7)~MeV positron energy range 

\begin{equation}
\label{eq2}
\chi^2 = \sum_{i=1}^N(R^{\mathrm{obs}}_i-k \times R_i^{\mathrm{pre}})^2/\sigma_i^2,
\end{equation}
where $R^{\mathrm{obs}}_i$ ($R^{\mathrm{pre}}_i$) is the observed (predicted) ratio of $\anti_particle\nu_e$ counting rates at the two detector positions and $\sigma_i$ is the statistical standard deviation of $R^{\mathrm{obs}}_i$, and $k$ is a normalization factor equal to the ratio of the total number of the IBD events per day at the bottom and top detector positions. The total numbers of IBD events per day at the two positions were equal in the calculations of the $R^{\mathrm{pre}}_i$. Thus, the $\chi^2$ does not depend on the integral IBD event rate dependence on the distance from the reactor core. Only differences in the positron energy shapes are considered.  This is the most conservative approach. The results do not depend even on the changes of the detector efficiency as long as they do not depend on the positron energy. This approach reduces also the sensitivity of the results to the position of the reactor fuel burning profile center and the reactor power. 

The middle detector position adds very little to the sensitivity to the sterile neutrino parameters since the distance between the middle and top (bottom) positions is twice smaller than that for the bottom and top positions. Inclusion of the middle/top ratio into the analysis makes it less transparent since it is correlated with the bottom/top ratio. Therefore, the data at the middle detector position were used only for the important crosschecks of the results. Figure~\ref{RatioMT} shows the ratio of the positron spectra at the middle and top detector positions. The limits on the sterile neutrino parameters from this ratio are fully consistent with the limits from the bottom and top position but they are considerably weaker. 

The oscillations due to the known neutrinos were neglected since at such short distances they do not change the $\anti_particle\nu_e$ spectrum in the studied energy range.
 The procedure was repeated for all points of the grid in order to get the whole exclusion area. 
Influence of systematic uncertainties in the parameters was estimated by repeating the analysis with different values of parameters. A point in the $\Delta m_{14}^2,~\sin^22\theta_{14}$ plane was included into the final excluded area if it appeared in the excluded areas for all tested variations of the parameters. The following variations of the parameters were tested:
\begin{itemize} 
\item The energy resolution multiplied by the factors 1.1 and 0.9 with respect to the MC predictions;
\item A flat background which gives $\pm0.1\%$ events at the top position of the detector which corresponds to 100\% variation of this background;
\item A background with the energy distribution identical to the distribution of the background produced by cosmic muons inside the detector. The fraction of such background was $\pm0.5\%$ of the IBD rate at the top position of the detector which corresponded to $\pm 15\%$ variation of this background;
\item The energy scale changed by $\pm2\%$;
\item All possible combinations of changes listed above;
\item The reduced range of the energies used in the fit to (1.5---6)~MeV.
\end{itemize}

\begin{figure}[h]
\begin{center}
\includegraphics[width=0.45\textwidth]{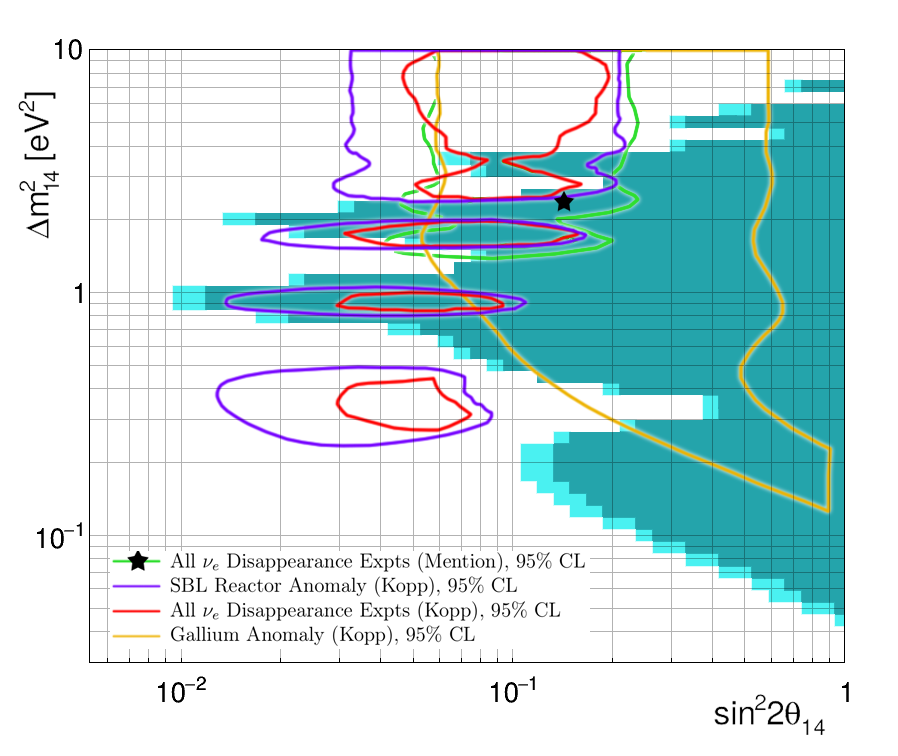}
\end{center}
\caption{\label{Exclusion}90\% (cyan) and 95\% (dark cyan) CL exclusion area in $\Delta m_{14}^2,~\sin^22\theta_{14}$ parameter space. The shaded area represents our analysis. Curves show allowed regions from neutrino disappearance experiments\cite{Mention2011,contours}, and the star is the best point from the RAA and GA fit\cite{Mention2011}.}
\end{figure}

Figure~\ref{Exclusion} shows the obtained 90\% and 95\% CL excluded area in the $\Delta m_{14}^2,\sin^22\theta_{14}$ plane. 
For some values of $\Delta m^2_{14}$ the obtained limits are more stringent than previous results\cite{Bugey,DBoscillations,NEOS}. It is important to stress that our results are based only on the comparison of the shapes of the positron energy distributions at the two distances from the reactor core measured with the same detector.
Therefore the results do not depend on the $\anti_particle\nu_e$ spectrum shape and normalization as well as on the detector efficiency.  The excluded area covers a large fraction of regions indicated by the GA and RAA. %In particular, the most preferred point $\Delta m_{14}^2=2.3~\rm{eV}^2,~\sin^22\theta_{14} =0.14$\cite{Mention2011} is excluded at more than 5$\sigma$ CL. 
In our analysis the point $\Delta m_{14}^2=1.4~\rm{eV}^2,~\sin^22\theta_{14} =0.05$ has the smallest $\chi^2 = 21.9$. The difference in $\chi^2$ with the 3$\nu$ case is 13.1. The significance of this difference will be studied taking into account systematic uncertainties after collection of more data this year.

\section*{Acknowledgments}
The authors are grateful to the directorates of ITEP and JINR for constant support of this work. 
The authors appreciate the permanent assistance of the KNPP administration and Radiation and Nuclear Safety Departments. 
The detector construction was supported by the Russian State Corporation ROSATOM (state contracts H.4x.44.90.13.1119 and H.4x.44.9B.16.1006). 
The operation and data analysis became possible due to the valuable support from the Russian Science Foundation grant 17-12-01145.

\begin{table}[tp]
 \caption{Ratio of the positron spectra at the bottom and middle detector positions to the spectrum at the top detector position (statistical errors only).}
 {\small
 \label{2}
\begin{tabular}{|c|cc|cc|}
\hline
%bin (MeV)& ratio & error\\
bin (MeV)& Bottom/Top & Error & Middle/Top & Error\\
\hline
1.00 -- 1.25 & 0.7175 & 0.0125 & 0.8616 & 0.0156\\
1.25 -- 1.50 & 0.7193 & 0.0102 & 0.8460 & 0.0127\\
1.50 -- 1.75 & 0.7272 & 0.0090 & 0.8566 & 0.0112\\
1.75 -- 2.00 & 0.7130 & 0.0081 & 0.8308 & 0.0101\\
2.00 -- 2.25 & 0.7114 & 0.0078 & 0.8398 & 0.0098\\
2.25 -- 2.50 & 0.7080 & 0.0076 & 0.8318 & 0.0095\\
2.50 -- 2.75 & 0.7108 & 0.0076 & 0.8407 & 0.0096\\
2.75 -- 3.00 & 0.7096 & 0.0077 & 0.8311 & 0.0096\\
3.00 -- 3.25 & 0.6915 & 0.0076 & 0.8237 & 0.0096\\
3.25 -- 3.50 & 0.7008 & 0.0080 & 0.8303 & 0.0101\\
3.50 -- 3.75 & 0.7138 & 0.0085 & 0.8362 & 0.0107\\
3.75 -- 4.00 & 0.6994 & 0.0088 & 0.8187 & 0.0110\\
4.00 -- 4.25 & 0.7044 & 0.0094 & 0.8298 & 0.0119\\
4.25 -- 4.50 & 0.6910 & 0.0100 & 0.8257 & 0.0128\\
4.50 -- 4.75 & 0.6889 & 0.0108 & 0.8283 & 0.0138\\
4.75 -- 5.00 & 0.7139 & 0.0123 & 0.8315 & 0.0154\\
5.00 -- 5.25 & 0.7147 & 0.0136 & 0.8098 & 0.0167\\
5.25 -- 5.50 & 0.7377 & 0.0160 & 0.8616 & 0.0201\\
5.50 -- 5.75 & 0.7244 & 0.0178 & 0.8172 & 0.0219\\
5.75 -- 6.00 & 0.7553 & 0.0212 & 0.8495 & 0.0259\\
6.00 -- 6.25 & 0.7483 & 0.0251 & 0.8573 & 0.0310\\
6.25 -- 6.50 & 0.6990 & 0.0275 & 0.7950 & 0.0340\\
6.50 -- 6.75 & 0.6907 & 0.0326 & 0.8096 & 0.0409\\
6.75 -- 7.00 & 0.6722 & 0.0393 & 0.8244 & 0.0509\\

\hline
\end{tabular}}
\end{table}

\section*{Appendix}

\label{appendix}

The Gaussian \cls method \cite{CLS} is a two-hypothesis test that compares in our case 
the three-neutrino (null) hypothesis (labeled $3\nu$) to an alternate
four-neutrino hypothesis (labeled $4\nu$). 
For each point in the $\Delta m_{14}^2,~\sin^22\theta_{14}$ plane we calculate the predictions for the positron spectra at the two positions in case of the $3\nu$ and $4\nu$ neutrino hypotheses. The calculations include the MC integration over the antineutrino production point in the reactor core and the positron position in the detector.
The distribution of the antineutrino production points in the reactor core was provided by the KNPP. We used the distribution averaged over the campaign.
It was checked that this approximation practically does not influence the final results. Then we convolute the resulting predictions with the detector resolution obtained by the MC for each energy point.
The large size of the reactor core and modest energy resolution lead to a substantial smearing of the oscillation pattern.
The theoretical predictions for $4\nu$ and $3\nu$ hypotheses  for the ratio of the positron spectra at the two distances at a given point in the $\Delta m_{14}^2,~\sin^22\theta_{14}$ plane were compared using the difference in $\chi^2$ for the two hypotheses $\Delta\chi^2_{\mathrm{exp}} = \chi^2_{4\nu} - \chi^2_{3\nu}$ (see Eqn.3). 
The difference in this $\chi^2$ has a Gaussian distribution with the mean value $\mu$ and the standard deviation $\sigma$ calculated using Asimov data set, a data sample with values following exactly theoretical curve for the corresponding $4\nu$ or $3\nu$ hypotheses and error bars taken from the real experiment\cite{CLS}. 
We calculate the corresponding $\Delta \chi^2_{\mathrm{Asimov}}$ by putting the experimental points with their statistical errors on the predicted curves for the $4\nu$ or $3\nu$ hypotheses respectively. The obtained $\Delta \chi^2_{\mathrm{Asimov}}$ determines the standard deviation of the $\Delta\chi^2_{\mathrm{exp}}$: $\sigma=\sqrt{|\Delta \chi^2_{\mathrm{Asimov}}|}$.
It is the same for the two Asimov data sets while the mean values $\Delta\chi^2_{\mathrm{Asimov}}(4\nu)$ and $\Delta\chi^2_{\mathrm{Asimov}}(3\nu)$ differ by the sign. Then we calculate the confidence levels for the $4\nu$ and $3\nu$ hypotheses by integration of the two Gaussian distributions with the obtained mean values $\pm\mu$ and the same standard deviation $\sigma$ from $\Delta\chi^2_{\mathrm{exp}}$ to infinity. %(Наверное лучше написать формулу). 
The $\cl_{4\nu}$ and $\cl_{3\nu}$ quantify the consistency of the data with the corresponding hypothesis and the \cls $= \cl_{4\nu}/\cl_{3\nu}$. The point in the $\Delta m_{14}^2,~\sin^22\theta_{14}$ plane is excluded at the 1-$\alpha$ confidence level if \cls $<\alpha$. Therefore the point is excluded only if the $3\nu$ hypothesis fits the data much better than the $4\nu$ hypothesis.
Hence only points for which the experiment has a sensitivity to distinguish the $4\nu$ and $3\nu$ hypotheses can be excluded. 
The systematic uncertainties are treated as the nuisance parameters in \cite{CLS}. The corresponding parameters with their errors are included into the minimization of the $\chi^2$ (see for example Eqn. 5\cite{CLS}).
We treat the systematic uncertainties differently. We repeat the \cls analysis without the nuisance parameters for all combinations of the systematic uncertainties taken at their maximal deviations from the nominal values. The systematic uncertainties in the energy resolution, energy scale, and the level of the two types of the background are taken into account as described in the section~\ref{data_analysis}. In addition, a separate study was performed to check the stability of the results with respect to the change of the energy range used in the analysis. It was checked that this approach is practically  identical (but more transparent) to using the systematic uncertainties as the nuisance parameters with the minimization of the $\chi^2$ over a limited set of deviations of the systematic parameters taken at their maximal values. Since the influence of the systematic uncertainties on the final results is small this approximation of the minimization procedure should be sufficient.
We can conclude that the \cls method used in this analysis provides the conservative exclusion limits on the sterile neutrino parameters.

%\section*{References}

\bibliography{danss_article.bbl}

\end{document}